\documentclass[10pt,journal]{IEEEtran}
\usepackage{cite}
\usepackage[pdftex]{graphicx}
\usepackage{amsmath,amssymb,amsfonts}
\usepackage{algorithmic}
\usepackage{algorithm}
\usepackage{booktabs}

\begin{document}
\title{Toward Packet Routing with Fully-distributed Multi-agent Deep Reinforcement Learning}
\author{Xinyu~You$^{\dagger}$,
        Xuanjie~Li$^{\dagger}$,
        Yuedong~Xu,
        Hui~Feng,
        Jin~Zhao,
        Huaicheng Yan

\IEEEcompsocitemizethanks{
\IEEEcompsocthanksitem X. You, X. Li, Y. Xu and H. Feng are with School of Information Science and Technology, Fudan University, Shanghai 200237, China. Emails: {\{xyyou18, xuanjieli16, ydxu, hfeng\}@fudan.edu.cn}
\IEEEcompsocthanksitem J. Zhao is with School of Computer Science, Fudan University, Shanghai 200237, China. Email: jzhao@fudan.edu.cn
\IEEEcompsocthanksitem H. Yan is with the Key Laboratory of Advanced Control and Optimization for Chemical Process of Ministry of Education, East China University of Science and Technology, Shanghai 200237, China, and also with the College of Mechatronics and Control Engineering, Hubei Normal University, Huangshi 435002, China. Email: hcyan@ecust.edu.cn.}
\thanks{$^{\dagger}$ The first two authors contributed equally to this work.}}

\markboth{Journal of \LaTeX\ Class Files,~Vol.~14, No.~8, August~2015}%
{Shell \MakeLowercase{\textit{et al.}}: Bare Demo of IEEEtran.cls for Computer Society Journals}

\IEEEtitleabstractindextext{%
\begin{abstract}
Packet routing is one of the fundamental problems in computer networks in which a router determines the next-hop of each packet in the queue to get it as quickly as possible to its destination. Reinforcement
learning (RL) has been introduced to design autonomous packet routing policies with local 
information of stochastic packet arrival and service. However, the curse of dimensionality of RL 
prohibits the more comprehensive representation of dynamic network states, thus limiting its potential benefit. 
In this paper, we propose a novel packet routing framework based on \emph{multi-agent} deep reinforcement learning (DRL) in which each router possess an \emph{independent} LSTM recurrent neural network for training and
decision making in a \emph{fully distributed} environment. The LSTM recurrent neural network extracts routing features from 
rich information regarding backlogged packets and past actions, and effectively approximates the value function 
of Q-learning. We further allow each route to communicate periodically with direct neighbors so that a broader 
view of network state can be incorporated. 
Experimental results manifest that our multi-agent DRL policy can strike the delicate balance between
congestion-aware and shortest routes, and significantly reduce the packet delivery time in 
general network topologies compared with its counterparts.

\end{abstract}

\begin{IEEEkeywords}
Packet routing, multi-agent learning, deep reinforcement learning, local communications
\end{IEEEkeywords}}

\maketitle
\IEEEdisplaynontitleabstractindextext
\IEEEpeerreviewmaketitle

\vspace{1.5cm}

\IEEEraisesectionheading{\section{Introduction}\label{sec:introduction}}
\IEEEPARstart
{P}{acket} routing is a very challenging problem in distributed and autonomous computer networks, especially 
in wireless networks in the absence of centralized or coordinated service providers. Each router decides to which neighbour it should send his packet in order to minimize the delivery time. The primary feature of packet routing resides in its fine-grained per-packet forwarding policy. No information regarding the network traffic is shared between neighbouring nodes. In contrast, exiting protocols use flooding approaches either to maintain a globally consistent routing table (e.g. DSDV \cite{DSDV}), or to construct an on-demand flow level routing table (e.g. AODV \cite{AODV}). The packet routing is essential to meet the dynamically changing traffic pattern in today's communication networks. Meanwhile, it symbolizes the difficulty of designing fully distributed forwarding policy that strikes a balance of choosing short paths and less congested paths through learning with local observations. 

Reinforcement learning (RL) is a bio-inspired machine learning approach that acquires knowledge by exploring the 
interaction with local environment without the need of external supervision \cite{RL_book}.  
Therefore, it is suitable to address the routing challenge in distributed networks where each node (interchangeable with 
router) measures the per-hop delivery delays as the reward of its actions and learns the best action accordingly. 
Authors in \cite{Q_routing} proposed the first multi-agent Q-learning approach for packet routing in a generalized network topology. 
This straightforward routing policy achieves much smaller mean delivery delay compared with the benchmark shortest-path approach. Xia et al. \cite{Xia} applied dual RL-based Q-routing approach to improve convergence rate of routing 
in cognitive radio networks. Lin and Schaar \cite{Joint_Q} adopted the joint Q-routing and power control policy for delay sensitive applications in wireless networks. More applications of RL-based routing algorithms can be found in \cite{RL_survey}. 
Owing to the well-known ``curse of dimensionality''\cite{curse}, the state-action space of RL is usually small such that the 
existing RL-based routing algorithms cannot take full advantage of the history of network traffic dynamics and cannot 
explore sufficiently more trajectories before deciding the packet forwarding. The complexity of training RL with large 
state-action space becomes an obstacle of deploying RL-based packet routing. 

The breakthrough of deep reinforcement learning (DRL) provides a new opportunity to a good many RL-based networking applications 
that are previously perplexed by the prohibitive training burden. With deep neural network (DNN) as a powerful approximator of Q-table, the network designer can leverage its advantages from two aspects: 
(1) the neural network can take much more information as its inputs, enlarging the state-action space for better policy making; 
(2) the neutral network can automatically abstract invisible features from high-dimensional input data \cite{high_dim}, thus achieving an end-to-end decision making yet alleviating the handcrafted feature selection technique. 
Recent successful applications include cloud resource allocation \cite{deeprm}, adaptive bitrate video streaming \cite{pensieve}, cellular scheduling \cite{scheduling}. DRL is even used to generate routing policy in \cite{SDN} against the dynamic traffic pattern that is hardly predictable. However, authors in \cite{SDN} considers a centralized routing policy that requires the global topology and the 
global traffic demand matrix, and operates at the flow-level. Inspired by the power of DRL and in view of the limitations of Q-routing\cite{Q_routing}, 
we aim to make an early attempt to develop fully-distributed packet routing policies using multi-agent deep reinforcement learning. 

In this paper, we propose a novel
multi-agent DRL algorithm named Deep Q-routing with Communication (DQRC) for fully distributed packet routing. 
Each router uses a carefully designed LSTM recurrent neural network (RNN)
to learn and infer the dynamic routing policy independently. 
DQRC takes the high-dimensional information as its input: 
the destinations of head-of-line (HOL) as well as 
backlogged packets, the action history,  
and the queue length of neighboring routers that are 
reported periodically. The action of DQRC is the next-hop of 
the HOL packet, and the packet delivery time is chosen as the reward of Q-learning so as to train the LSTM neural network 
at each router. The intuitions of our design 
stand for twofold relation to the queueing process. 
On one hand, the action history is closely related to the congestion of next hops, the number of backlogged packets  indicates the load of the current router, and knowing the destinations of outgoing packets avoids pumping them into the same adjacent routers. 
On the other hand, with a lightweight communication scheme, an agent can acquire the queue information of its direct neighbors and learn to send packets to less congested next hops. With such a large input space, existing RL approaches such Q-routing \cite{Q_routing} cannot handle the training online so that
the training of deep neural networks using RL rewards becomes a necessity. 
DQRC is fully distributed in the sense that each router is configured with an independent neural network for parameter update and decision making. This differs from the recent multi-agent DRL learning framework in other domains \cite{DQN_routing} where the training of neural networks are simultaneous and globally consistent. The training of multi-agent DRL is usually difficult (e.g. convergence and training speed), while DQRC proves the feasibility of deploying DRL-based packet routing in the dynamic environment.

With Q-routing \cite{Q_routing} and Backpressure \cite{Backpressure} as benchmarks, our experimental results reveal a few interesting observations. 
Firstly, DQRC significantly outperforms the other two representative algorithms in terms of the average delivery delay with different network loads and topologies. With careful examination of DQRC routing policy, we observe that each router makes adaptive routing decision by considering more information than the destination of the HOL packet, thus avoiding congestion on ``popular'' paths. Secondly, the lightweight communication mechanism is very beneficial to DQRC. An agent only shares 
its raw queue length other than a host of DNN parameters 
with its neighbors, and this sharing can be infrequent or delayed for several time slots with merely gentle increase of delivery time. Thirdly, DQRC is robust to the hyper-parameters of LSTM neural network, indicating that a moderate complexity of neural networks 
(e.g. 3 hidden layers and 128 neurons in a hidden layer) is sufficient.

The remainder of this paper is organized as follows: Section \ref{sec:background} reviews the background knowledge of RL, DRL and POMDP. Section \ref{sec:design} presents our design of DQRC. The delivery delay of the proposed algorithm are evaluated in Section \ref{sec:evaluation} with Q-routing and Backpressure as the benchmark. Section \ref{sec:discussion} is devoted to making discussions about future study and challenges. Section \ref{sec:conclusion} concludes this work.

\section{Background and Literature}
\label{sec:background}
In this section, we briefly review the traditional routing algorithms, RL and DRL techniques and their applications to routing problem. We then put forward the necessity of fully-distributed learning for real-world routing problem. Finally, the background of POMDP is included.

\subsection{Traditional Routing Algorithm}

As a profound and globally applicable routing algorithm, shortest-path algorithm reveals how to transfer all the data as quickly as possible. Within all the algorithms in shortest-path class, Bellman-Ford algorithm \cite{Bellman} and Dijkstra's algorithm \cite{Dijkstra} are crucial to the development of network protocols. The definition of a shortest path may vary with different contexts, such as transmission delay or number of hops in the network. It is easy to understand that along the shortest path between two nodes, data packets can be delivered costing the least amount of time provided that there is no congestion along the route. However, these assumptions are unreachable for the realistic network.
When the network load is heavy, shortest-path algorithm will lead to severe backlogs in busy routers and thus necessitates manual monitoring and adjustment.

Backpressure \cite{Backpressure} has shown its efficiency in dynamic traffic routing in multi-hop network by using congestion gradients. Each node maintains different queues for all of the potential destination nodes and each packet is assigned to one queue according to its destination. The basic idea of Backpressure is to utilize the differential of the queue backlogs between the current node and its neighbor nodes as the pressure to drive packet transmission. Backpressure has been widely studied in the literature because of its simplicity and asymptotic optimality at heavy traffic regime, but there are still some drawbacks hindering its wider application:
Backpressure may suffer from poor delay performance particularly in light load, in which case there is not enough pressure to push data towards the destination and therefore packets may choose unnecessary longer paths and even loops.

\subsection{RL Routing Algorithm}
\label{sec:RL}
Based on the mapping relationship between observed state and execution action, RL aims to construct an agent to maximize the expected discounted reward through the interaction with environment. Without prior knowledge of which state the environment would transition to or which actions yield the highest reward, the learner must discover the optimal policy by trial-and-error.

The first attempt to apply RL in the packet routing problem is Q-routing algorithm, which is a variant of Q-learning \cite{RL_book}. Since Q-routing is essentially based on multi-agent approach, each node is viewed as an independent agent and endowed with a Q-table to restore Q-values as the estimate of the transmission time between that node and the others. With the aim of shortening average packet delivery time, agents will update their Q-table and learn the optimal routing policy through the feedback from their neighboring nodes when receiving the packet sent to them. Despite the superior performance over shortest-path algorithm in dynamic network environment, Q-routing suffers from the inability to fine-tune routing policy under heavy network load and the inadequate adaptability of network load change. To address these problems, other improved algorithms have been proposed such as PQ-routing \cite{Pred_Q} which uses previous routing memory to predict the traffic trend and DRQ-routing \cite{Dual_Q} which utilizes the information from both forward and backward exploration to make better decisions. 

\subsection{DRL Routing Algorithm}
DRL embraces the advantage of deep neural networks \cite{rbf1} \cite{rbf2} to the training process, thereby improving the learning speed and the performance of RL \cite{survey_DRL}. One popular algorithm of DRL is Deep Q-Learning (DQL) \cite{DQN}, which implements a Deep Q-Network (DQN) instead of Q-table to derive an approximation of Q-value with special mechanisms of experience replay and target Q-network. 

Recently, network routing problems are solved using DRL under different environment and optimization targets. Based on the control model of the agent, these algorithms can be categorized as follows:

\noindent\textbf{Class 1: Single-agent learning.}

 Single-agent algorithm treats the network controller as a central agent which can observe the global information of the network and control the packet transmission at each router. Both the learning and execution process of this kind of algorithm are centralized \cite{commu}, in other words, the communication between routers are not restricted during training and execution.

SDN-Routing \cite{SDN} presents the first attempt to apply single-agent DRL in the routing optimization of traffic engineering. The traffic demand which represents the bandwidth request between each source-destination pair is viewed as the environment state. The network controller determines the transmission path of packets to achieve the objective of minimizing the network delay. Another algorithm \cite{Learning_to_route} considers a similar network model while taking minimum link utilization as the optimization target.

\noindent\textbf{Class 2: Multi-agent learning.}

Cooperative control in multi-agent systems provides an alternative way to solve a complicated problem that are hard to be performed for one agent \cite{linear} \cite{non-linear}.
In multi-agent learning, each router in the network is treated as a single agent which can observe partial environment information and take actions according to its own routing policy.

The first multi-agent DRL routing algorithm is DQN-routing \cite{DQN_routing} that is the combination Q-routing and DQN. Each router is regarded as an agent whose parameters are shared by each other and updated at the same time during training process (centralized training), but it provides independent instructions for packet transmission (decentralized execution). The comparison with contemporary routing algorithms confirms a substantial performance gain.

Nevertheless, algorithms with centralized learning process stated above are hard to be applied in the realistic network. The centralized learning controller is usually unable to gather environment transitions from widely distributed routers once an action is executed somewhere and to update the parameters of each neural network simultaneously caused by the limited bandwidth \cite{lb1}\cite{lb2}. 

Accordingly, for better application in real-world scenario, the routing algorithm we propose is executed in a fully-distributed way \cite{fully-distributed}, which means both the training process and the execution process are decentralized. Under these circumstances, each agent owns its unique neural network with independent parameters for policy update and decision making, thereby avoiding the necessity for the communications among routers in the process of environment transition collection and parameter update. 

\subsection{POMDP}
In the real-world environment, it is rare that agents can observe the full state of the system. This type of problem can be modelled as partially observable Markov decision processes (POMDPs) where the observation agents receive is only the partial glimpse of the underlying system state \cite{DRQN}. 
The partial observation of system state will aggravate the non-stationarity problem in multi-agent reinforcement learning (MARL).

To tackle with this problem, an efficient solution is to utilize Deep Recurrent Q-Network (DRQN) \cite{DRQN} which is capable of capturing invisible features based on time-continuous observations.
Derived from DQN, this ingenious neural network architecture substitutes some fully-connected layers by Long Short Term Memory (LSTM) \cite{LSTM_1997} layers which can not only store and reuse history information to explore temporal features, but also split inputs into several parts to search connections between them. The LSTM layer contains special units called memory blocks in which memory cells can store the temporal state of the network and gates are used to control the flow of information \cite{LSTM}. The existence of LSTM helps alleviate the dilemma that the present task cannot receive the relevant information long before it. 

Another promising direction is to design a cooperation or communication mechanism to help agents learn from not only its own knowledge but also public information broadcast from other agents. Two MARL algorithms are put forward in \cite{commu} to address partially observable multi-agent decision making problems: reinforced inter-agent learning (RIAL) and differentiable inter-agent learning (DIAL). The former uses independent DQN, while the latter backpropagates error derivatives through communication channels during learning. With centralised learning but decentralised execution, the communication between agents promise the success in MARL domain.

Inspired by these ideas, we propose our algorithm DQRC with fully distributed multi-agent deep reinforcement learning. DQRC amalgamates traditional DQN with LSTM architecture as well as communication mechanism to address POMDPs in packet routing problem.

\section{Design}
\label{sec:design}
We establish the mathematical model of the packet routing problem and describe the representation of each element in the reinforcement learning formulation.
We then propose a novel deep recurrent neural network architecture and the corresponding training algorithm.

\begin{figure}[t]
	\centerline{\includegraphics[scale=0.8]{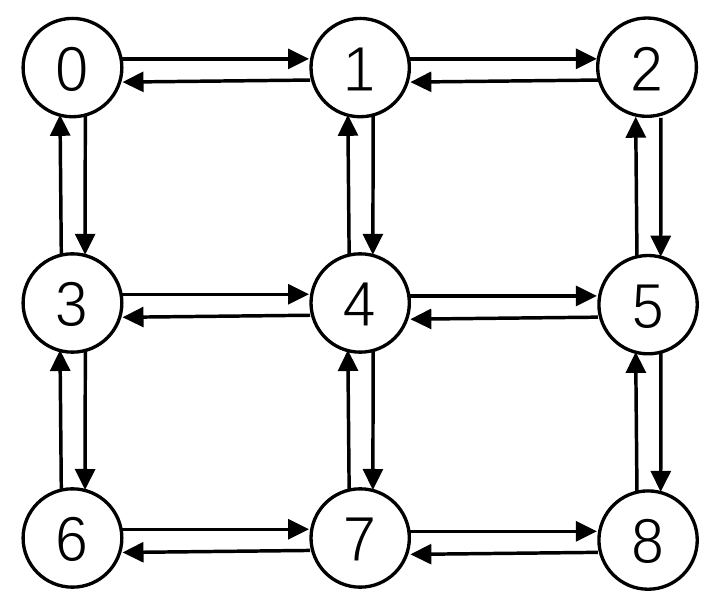}}
	\caption{3x3 network topology.}
	\label{fig:topology}
	\vspace{-0cm}
\end{figure}

\subsection{Mathematical Model}
We consider a network consisting of routers and links connected between them, and packet routing is the transmission of packets from a source to its destination through intermediate nodes. We now present the the mathematical model of the packet routing problem.

\textbf{Network.} The network is modeled as a directed graph $\mathcal{G=(N, E)}$, where $\mathcal{N}$ and $\mathcal{E}$ are defined as finite sets of nodes and transmission links between them respectively. A simple network topology can be found in Fig. \ref{fig:topology}, containing 9 nodes and 12 pairs of bidirectional links. Each packet is originated from node $s$ and destined for node $d$: $s, d \in \mathcal{N}$ and $s\neq d$ with randomly generated intervals.

\textbf{Routing.} The mission of packet routing is to transfer each packet to its destination through the relaying of multiple routers. The queue of routers follows the first-in first-out (FIFO) criterion. Each router $n$ constantly delivers the HOL packet to its neighbor node $v$ until that packet reaches its termination. 

\textbf{Target.} The packet routing problem aims at finding the optimal transmission path between source and destination nodes based on some routing metric, which, in our experiment, is defined as the average delivery time of packets. Formally, we denote the packet set as $\mathcal{P}$ and the total transmission time as $t_{p}$  for each packet $p: p \in \mathcal{P}$. Our target is to minimize the average delivery time $T = \sum_{p \in \mathcal{P}} t_{p} / K$, where $K$ denotes the number of packets in $\mathcal{P}$.

\subsection{Reinforcement Learning Formulation}
\label{sec:RL_formulation}
{Packet routing can be modeled as a multi-agent reinforcement learning problem with partially observable Markov decision processes (POMDPs) \cite{POMDP}. Each node is an independent agent and learns its routing policy by observing the local network state and communicating with its neighbor nodes.
Therefore, we will describe the definitions of each element in reinforcement learning for a single agent.}

\textbf{State space.} The packet $p$ to be sent by agent $n$ is defined as \textit{current packet}. We denote the state space of agent $n$ as $S_{n}: \{d_{p}, E_{n}, C_{n}\}$, where $d_{p}$ is the destination of the current packet, $E_{n}$ is some extra information related to agent $n$, $C_{n}$ is the information shared form the neighbor nodes of agent $n$. At different time steps, the state observed by the agent is time varying due to the dynamic change of network traffic.

\textbf{Action space.} The action space of agent $n$ is defined as $A_{n}: \mathcal{V}_{n}$, where $\mathcal{V}_{n}$ is the set of neighbor nodes of node $n$. Accordingly, for each agent, the size of action space equals to the number of its adjacent nodes. For example, node 4 in Fig.\ref{fig:topology} has four candidate actions. Once a packet arrives at the head of queue at time step $t$, agent $n$ observes the current state $s_{t} \in S_{n}$ and picks an action $a_{t} \in A_{n}$, and then the current packet is delivered to the corresponding neighbor of node $n$.

\textbf{Reward.} We craft the reward to guide the agent towards effective policy for our target: minimizing the average delivery time. The reward at time step $t$ is set to be the sum of queueing time and transmission time: $r_{t}=q+l$, where the former $q$ is the time spent in the queue of agent $n$, and the latter $l$ is referred to as the transmission latency to the next hop.

\begin{figure}[t]
    \centerline{\includegraphics[scale=0.45]{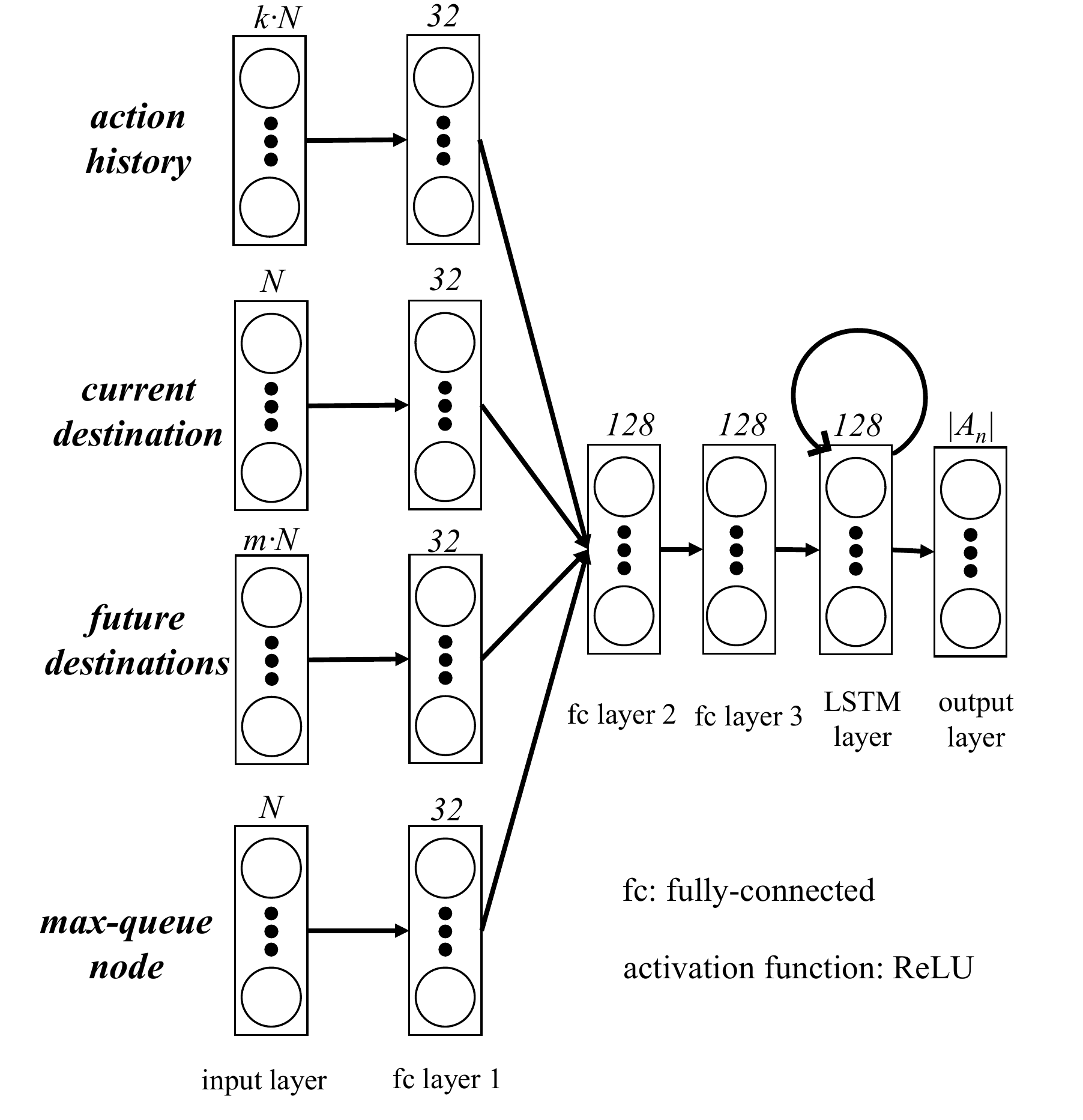}}
    \caption{LSTM neural network with ReLU activation.}
    \label{fig:lstm_network}
    \vspace{-0cm}
\end{figure}

\subsection{Deep Neural Network}
\label{sec:dnn}

We will introduce the architecture of the neural network designed for our algorithm in this part. Note that, in the formulation of fully-distributed reinforcement learning, each node is an individual agent and therefore possesses its own neural network for decision making. Accordingly, the following description of the neural network architecture is tailored for a single agent.

As shown in Fig.\ref{fig:lstm_network}, we build a deep recurrent neural network with three fully-connected layers and one LSTM layer. The input of the neural network can be classified into four parts: 
\begin{itemize}
  \item [(1)] 
  current destination: the destination node of the current packet.     
  \item [(2)]
  action history: the executed actions for the past $k$ packets sent out just before the current packet.
  \item [(3)]
  future destinations: the destination nodes of the next $m$ packets waiting behind the current packet.
  \item [(4)]
  max-queue node: the node which has the longest queue among all the neighbor nodes of the current node. 
\end{itemize}
In the formulation of the state space in Section \ref{sec:RL_formulation}, the current node corresponds to $d_{p}$, the action history and the future destinations corresponds to $E_{n}$, the max-queue node corresponds to $C_{n}$.

Before being input into the neural network, all of the above information will be processed with one-hot encoding. Take the first part of input information as an example, if the current packet is destined to node 4, the one-hot encoding result of current destination is [000010000]. Therefore, the the number of input neurons equals to $(1+k+m+1)\times N$, where $N$ is defined as the total number of nodes in the network topology.

The first hidden layer is a concatenation of four subsets of hidden neurons. Each subset possesses 32 neurons and is fully connected with the corresponding part of the input neurons independently. Following the first hidden layer, another two hidden layers with 128 neurons are added.
 
In partially observable environment, each agent can only receive an observation $s$ which is correlated with the full environment state. Inspired by Deep  Recurrent Q-networks (DRQN)\cite{DRQN}, we add a LSTM layer to maintain an internal state and aggregate observations over time. 
Despite the partial observation $s$, the hidden state of the agent $h$ will be included to represent Q-value which is defined as $Q(s,h,a)$.

Furthermore, the size of the output layer and the agent’s action space $|A_{n}|$ are identical, and the value of each output neuron is the estimated Q-value of the corresponding action. With this change of the representation for Q-value, we try to update the parameter of neural networks instead of the value of the Q-table. 

We use Rectified Linear Unit (ReLU) as the activation function and Root Mean Square Prop (RMSProp) as the optimization algorithm.

\subsection{Learning Algorithm}
By integrating Q-routing and DRQN, we propose the packet routing algorithm with multi-agent deep reinforcement learning, where both training and execution process are set decentralized.
The pseudo-code of the learning algorithm, which we call Deep Q-routing with Communication (DQRC), is shown in Algorithm 1, in which the initialization and the training process are identical for each node. 

Each node $i$ is treated as an individual agent and possesses its own neural network $Q_{i}$ with particular parameter $\theta_{i}$ to estimate the state-action value function  $Q_{i}(s,h,a;\theta_{i})$, which represents the expected delivery time for a packet to reach the destination when the agent executes action $a$ in state $s$ and hidden state $h$. Replay memory $D_{i}$ with capacity of 100 is also initialized independently for each agent to restore its environment transitions, and from it a random min-batch with size of 16 will be sampled for the update of its network parameters.

\begin{algorithm}[t]
	\caption{Deep Q-routing with Communication (DQRC)}
	\label{alg:A1_2}
	\begin{algorithmic}
		\STATE // initialization
		\FOR {agent $i$ = 1, $N$}
		\STATE {Initialize replay buffer $D_{i}\gets\varnothing$}
		\STATE {Initialize Q-network $Q_{i}$ with random weights $\theta_{i}$} 
		\ENDFOR
		
		\STATE
		\STATE // training process
		\FOR {episode = 1, $M$}
		\FOR {each decision epoch $t$}
		\STATE Assign current agent $n$ and packet $p$
		\STATE Observe local information $d_{p}$ and $E_{n}$
		\STATE Collect shared information $C_{n}$
		\STATE Integrate current state $s_{t}:\{d_{p},E_{n},C_{n}\}$ and hidden state $h_{t}$
		\STATE Select and execute action $$a_{t}=
		\begin{cases}
		\text{a random action}&\text{with probability $\epsilon$}\\
		argmax_{a} Q_{n}(\theta_{n})&\text{with probability $1-\epsilon$}
		\end{cases}$$
		\STATE Forward $p$ to next agent $v_{t}$
		\STATE Calculate and collect reward $r_{t}$
		\STATE Observe next state $s_{t+1}$ and next hidden state $h_{t+1}$
		\STATE Set transmission flag $f_{t}=
		\begin{cases}
		1& \text{$v_{t} = d_{p}$}\\
		0& \text{otherwise}
		\end{cases}$
		\STATE Store transition $(s_{t}, h_{t}, r_{t}, v_{t}, s_{t+1}, h_{t+1}, f_{t})$ in $D_{n}$
		\STATE Sample a random batch $(s_{j}, h_{j}, r_{j}, v_{j}, s_{j+1}, h_{j+1}, f_{j})$ from $D_{n}$
		\STATE Set $y_{j} = r_{j}  + max_{a'} Q_{v_{j}} (s_{j+1} , h_{j+1}, a' ; \theta_{v_{j}})(1- f_{j})$
		\STATE  $\theta_{n}\gets$ \textit{GradientDescent} ( $(y_{j} - Q_{n} (s_{j} , h_{j}, a_{j}; \theta_{n}))^{2}$ )
		\ENDFOR
		\ENDFOR
	\end{algorithmic}
\end{algorithm}

For each decision epoch $t$ when a packet $p$ arrives at the head of line of a certain node $n$, agent $n$ will observe local information $d_{p}$ and $E_{n}$ and collect shared information $C_{n}$ through the communication with neighbor nodes. By integrating current state $s_{t}:\{d_{p},E_{n},C_{n}\}$ and hidden state $h_{t}$, agent $n$ will execute an action $a_{t}$ based on $\epsilon$-greedy policy, which means agent $n$ will choose a random action from its action space $A_{n}$ with probability $\epsilon$ or choose the action with the highest Q-value with probability $1-\epsilon$. The assignment of $a_{t}$ is given by:
$$a_{t}=
\begin{cases}
\text{a random action}&\text{with probability $\epsilon$}\\
argmax_{a} Q_{n}(s_{t},h_{t},a_{t};\theta_{n})&\text{with probability $1-\epsilon$}
\end{cases}\eqno(3.1)$$
Then the current packet $p$ is forwarded to the corresponding neighbor node $v_{t}$, and the reward $r_{t}$ is calculated and sent back to agent $n$.
Current state and hidden state will transition to $s_{t+1}$ and $h_{t+1}$ respectively. Besides, the transmission flag $f_{t}$ will be set to 1 if the next node $v_{t}$ matches the packet’s destination $d_{p}$ or set to 0 otherwise. The assignment of $f_{t}$ is given by:
$$f_{t}=
\begin{cases}
1& \text{$v_{t} = d_{p}$}\\
0& \text{otherwise}
\end{cases}\eqno(3.2)$$
After the feedback of these information, agent $n$ will record this transition $(s_{t}, h_{t}, r_{t}, v_{t}, s_{t+1}, h_{t+1}, f_{t})$ into its replay memory $D_{n}$. Different from the sequential update process of DRQN, a training batch $(s_{j}, h_{j}, r_{j}, v_{j}, s_{j+1}, h_{j+1}, f_{j})$ is sampled randomly from $D_{n}$ to avoid violating the DQN random sampling policy. 
As a result of the unstable environment caused by the multi-agent characteristic, the remaining delivery time $\tau$ that packet $p$ is expected to spend from $v_{t}$ to $d_{p}$  need to be recalculated before the training process. $\tau$ is given by:
$$\tau=max_{a'} Q_{v_{j}} (s_{j+1}, h_{j+1}, a' ; \theta_{v_{j}})\eqno(3.3)$$
At the end of the decision epoch $t$, the method of gradient descent is used to fit the neural network $Q_{n}(\theta_{n})$. 
The target value $y_{j}$ is the sum of the immediate reward $r$ and the remaining delivery time $\tau$:
$$y_{j} = r_{j}  + \tau (1- f_{j})\eqno(3.4)$$
The parameter of $Q_{n}(\theta_{n})$ can be trained by minimising the loss function $L_{t}$:
$$L_{t} = (y_{j} - Q_{n} (s_{j} , h_{j}, a_{j}; \theta_{n}))^2\eqno(3.5)$$
After differentiating the loss function $L_{t}$ with respect to the weights $\theta_{n}$, we can update $\theta_{n}$ by:
$$\theta_{n}\gets \theta_{n} + \alpha\nabla_{\theta_{n}}(y_{j} - Q_{n} (s_{j} , h_{j}, a_{j}; \theta_{n}))^{2}\eqno(3.6)$$
where $\alpha$ is the learning rate.
In this way, the network parameters of each agent are updated with episodic training until convergence.

\section{Evaluation}
\label{sec:evaluation}
We conducted several experiments in the simulation environment of computer network to evaluate the performance of DQRC in both online and offline training mode. Our experiments cover a broad set of network conditions and topologies. Then we will give an explanation for the comparison results from the perspective of input information and learned policy. Furthermore, a deep dive into DQRC will be made to test its robustness to various parameters.

\subsection{Simulation Environment}
\label{sec:se}
We now describe the settings of our simulation environment and another two algorithms to be compared with DQRC.

\textbf{Topology.} The topology of the network we used is the same as Fig. \ref{fig:topology}, which remains static in the whole experiment. Despite the simple structure, we can explore new insights into packet routing, and actually a more complex network will lead to similar results. All the nodes and links in the network share the same attributes: the buffer size of each node is unlimited and the bandwidth of each link equals to the packet size, in which case only a single packet can be transmitted at a time.

\textbf{Packet.} A certain proportion, named \textit{distribution ratio}, of packets are generated from node 0 (busy ingress-router) to node 8  (busy egress-router), while the other packets’ source and destination are chosen uniformly. Packets are introduced into the network with the same size and 
their generated intervals are fixed at a certain value which is inversely proportional to the load of network traffic. 

\textbf{Time setting.} The time during the simulation is measured by milliseconds. The transmission time between adjacent nodes a packet has to spend is set to 1.0ms. The performance criterion of the experiment is the average delivery time of packets within a certain period.

\textbf{Compared algorithms.}
Both Q-routing and Backpressure algorithms stated in Section \ref{sec:background} are included for performance comparison. Since \cite{Q_routing} has proved the inferior performance of shortest-path algorithm, it will not be taken into comparison.
In our experiment, in order to follow the FIFO criterion, the traditional Backpressure algorithm is tailored to route the HOL packet to the neighbor node with the minimal number of packets with the same destination.

\begin{figure}[t]
	\centerline{\includegraphics[scale=0.4]{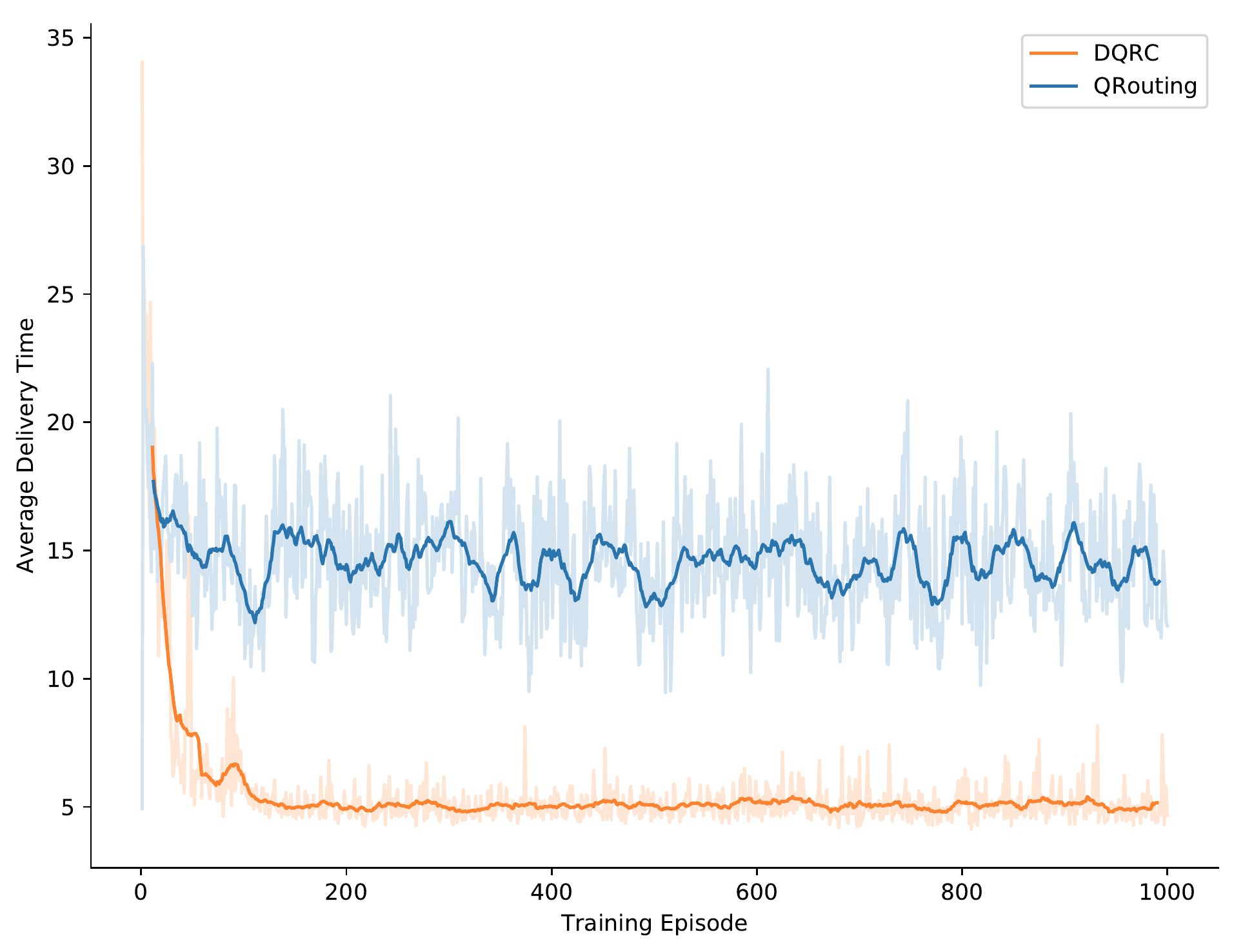}}
	\caption{Offline training speed comparison.}
	\label{fig:training}
	\vspace{-0cm}
\end{figure}

\subsection{Offline-training Online-test Result}
\label{sec:experiment}

In offline experiments, we generated a fixed packet series containing 1000 packets as the training set on which we trained the neural network and Q-table separately. Instead of ending with all the packets routed to their destinations, each training episode stops in 100ms, in order to meet the characteristic of endless packet in online environment. After 1000 training episodes, we restored the well-trained models to compare their performance in a new test environment where packets were generated at the corresponding network load level but different source-destination pairs from the training set. Note that Backpressure does not need training and can be applied to test directly.

\textbf{Training speed.} 
With the fixed packet sequence whose generated interval is 0.5ms and distribution ratio is 70\%, we trained DQRC and Q-routing and compared their training speeds. Fig. \ref{fig:training} shows the variation trend of average packet delivery time along with the training episode. The solid line is the smoothed result of the true value depicted in the shadowed part.  
Though performing worse in the first few episodes, DQRC convergences quickly and keeps stable at a lower delay level. On the contrary, Q-routing fluctuates violently from time to time and never converges.
From the perspective of either convergence speed or final training result, DQRC outperforms Q-routing in our simulation.

\begin{figure}[t]
	\centerline{\includegraphics[scale=0.45]{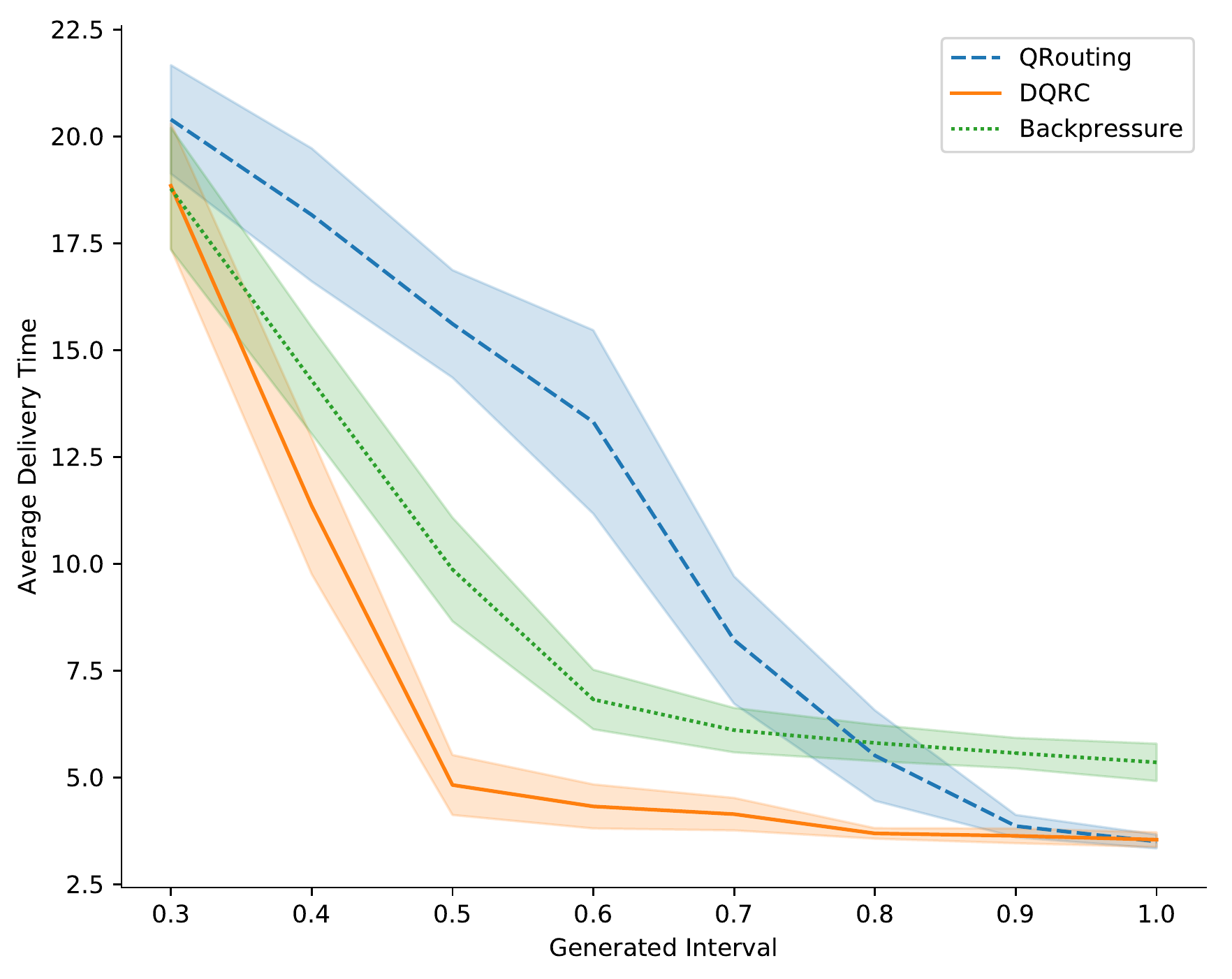}}
	\caption{Offline test with different network loads.}
	\label{fig:gi}
	\vspace{-0cm}
\end{figure}

\textbf{Network load.}
In terms of the distribution ratio at 70\%, we carried out online tests with various packet generated intervals ranging from 0.3ms to 1.0ms and recorded 50 sets of result for the first 100ms under each parameter. The average value and standard deviation of these resulting data are depicted with curve and shadow separately in Fig. \ref{fig:gi}. As expected, we can clearly see that:
\begin{itemize}
  \item [(1)] 
  All three algorithms present a rising tendency on average delivery time as the load becomes heavier. 
  DQRC outperforms the other two algorithms at all network load levels.
  \item [(2)]
  Q-routing behave equivalently with DQRC when the generated interval is between 0.9ms and 1.0ms (low network load). Conversely, Backpressure has fairly good performance when when the generated interval is between 0.4ms and 0.8ms (high network load). This interesting phenomenon may reveal the fact that DQRC has the merits of Q-routing and Backpressure while discarding their shortcomings.
  \item [(3)]
  The width of shadow band of DQRC is the narrowest, meaning that our proposed algorithm performs more stable during the test and is robust to randomly generated packet set.
\end{itemize}

\begin{figure}[t]
	\centerline{\includegraphics[scale=0.45]{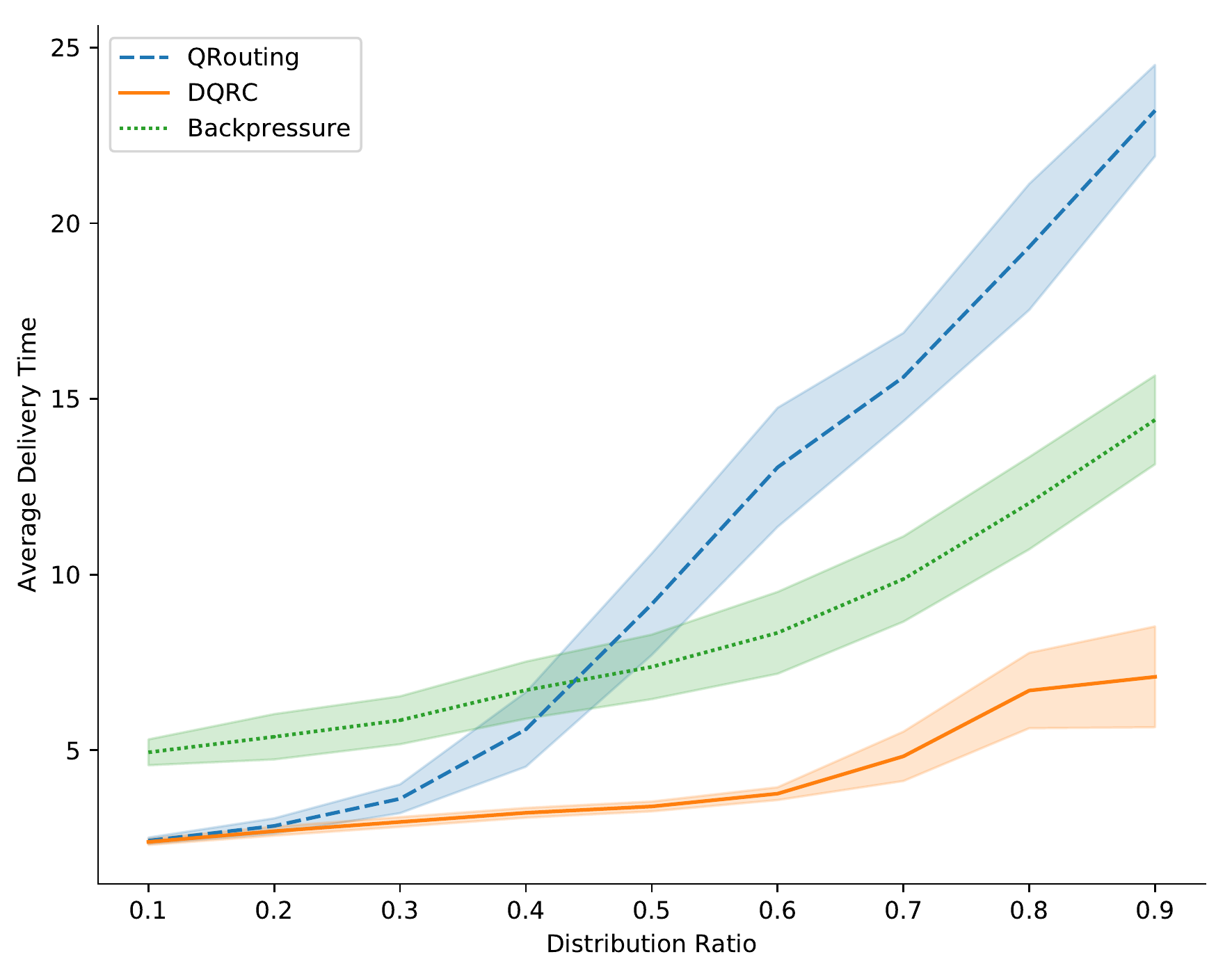}}
	\caption{Offline test with different distribution ratios.}
	\label{fig:dr}
	\vspace{-0.0cm}
\end{figure}

\textbf{Distribution ratio.} 
We collected another 50 sets of results from online tests by adjusting the distribution ratio from 10\% to 90\% and fixing the generated interval at 0.5ms. Similarly, each test ends in 100ms. From the result diagram shown in Fig. \ref{fig:dr}, we can obtain more appealing discoveries:
\begin{itemize}
  \item [(1)] 
  As the distribution ratio continues increasing, the average delivery time of all three algorithms is prolonged because larger distribution ratio indicates more pressure on the links connnetting node 0 and node 8. 
  \item [(2)]
  The distribution ratio for Q-routing and Backpressure with good performance is 10\%--40\% and 50\%--90\% respectively, but DQRC works well in either case. This result draws the same conclusion that DQRC possesses the features of Q-routing and Backpressure as that stated in the above part.
  \item [(3)]
  Considering DQRC, there is almost no shadow area around the solid line before the distribution ratio reaches 60\%, indicating the good robustness to the randomness of packet generation.
  \item [(4)]
  The slope of Q-routing is the largest, showing that Q-routing has higher sensitivity to the change of distribution ratio than the other two algorithms. By contrary, DQRC and Backpressure are more robust to this kind of traffic change.
\end{itemize}

 In summary, no matter in which kind of test environment, DQRC integrates the merits of both Q-routing and Backpressure and performs better in the following three aspects: 
 \begin{itemize}
  \item [(1)] 
  The average delivery time of DQRC is the shortest regardless of generated interval or distribution ratio.
  \item [(2)]
  DQRC is more adaptable and resistant to the change of network load and spatial distribution of packet generation.
  \item [(3)]
  DQRC frees itself from random errors caused by the network disturbances.
\end{itemize}

\subsection{Online-training Online-test Result}
\label{sec:online}
In online simulation environment, packets are generated all the time following the regulations described in Section \ref{sec:se}. The simulation timeline is split into intervals of 100ms and for each interval the average delivery time of transmitted packets is recorded. In order to test each algorithm's adaptability to the change of traffic pattern, we initially set the generated interval of packets to 1.0ms and suddenly change it to 0.7ms at time 4000ms and reset it to 1.0ms at time 8000ms. Before put into online test, DQRC and Q-routing are well-trained in offline training environment with generated interval set to 1.0ms, and during the simulation the parameters of neural networks and the value of the Q-table are updated from time to time. Note that Backpressure does not need training.

\begin{figure}[t]
	\centerline{\includegraphics[scale=0.45]{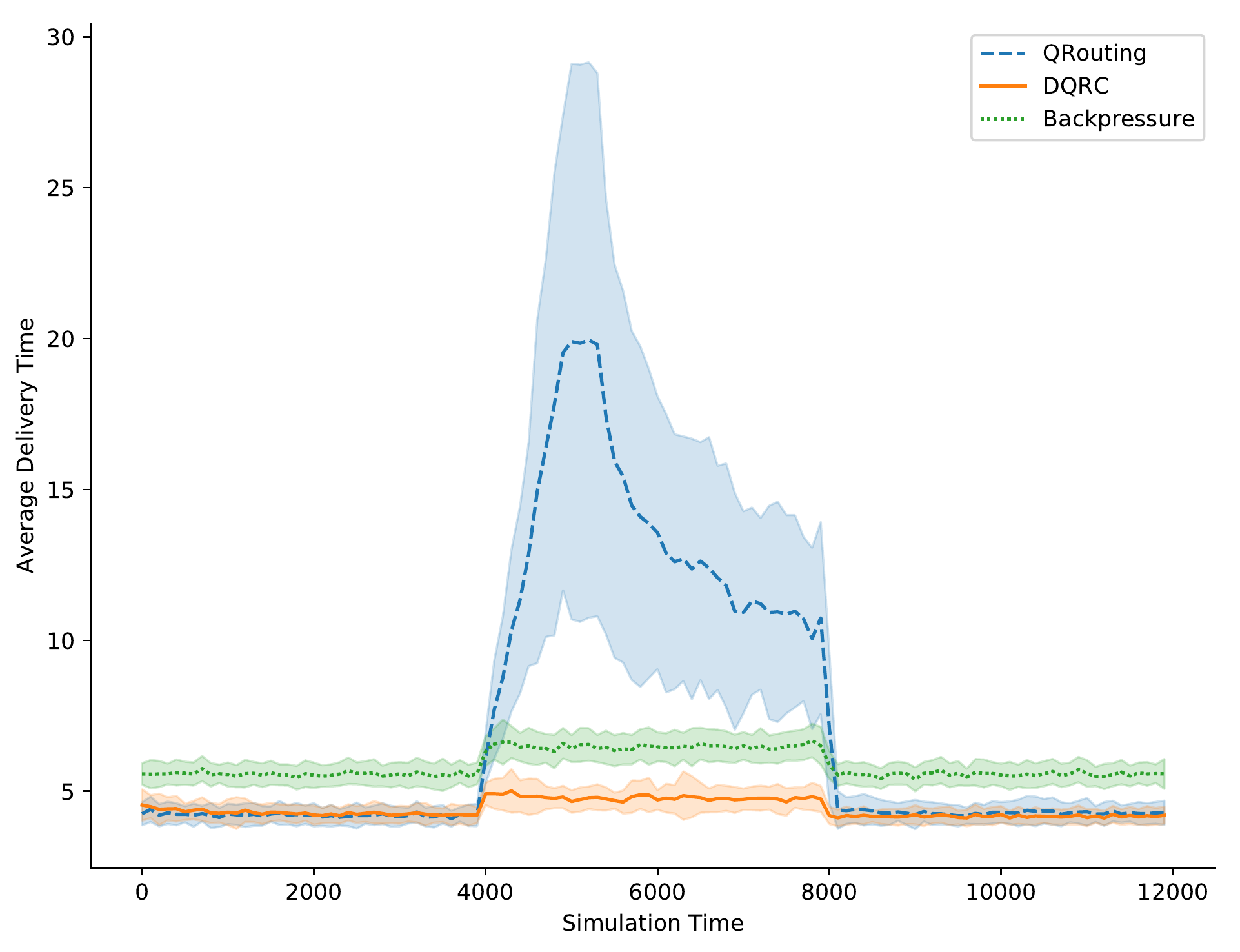}}
	\caption{Online test result with changing network loads.}
	\label{fig:online}
	\vspace{-0cm}
\end{figure}

Fig. \ref{fig:online} plots the average result of 50 online tests with different packets sets. 
We can clearly see that (1) Q-routing have comparable performance with DQRC in the first and the third stage. However, when the network load begin to increase at 4000ms, Q-routing find it hard to converge  to a new routing policy to address the growing number of packets, leading to larger transmission delay and unstable performance; (2) Though behaving fairly well in the second stage, Backpressure has an inferior performance compared to DQRC and Q-routing in lightly loaded network due to insufficient pressure to push packets to their destinations; (3) DQRC achieves the shortest delivery time and the  lowest fluctuations in all three stages. Besides, DQRC converges more quickly than Q-routing, and therefore is more adaptable to dynamic changes of network load.

\begin{figure}[t]
	\centerline{\includegraphics[scale=0.7]{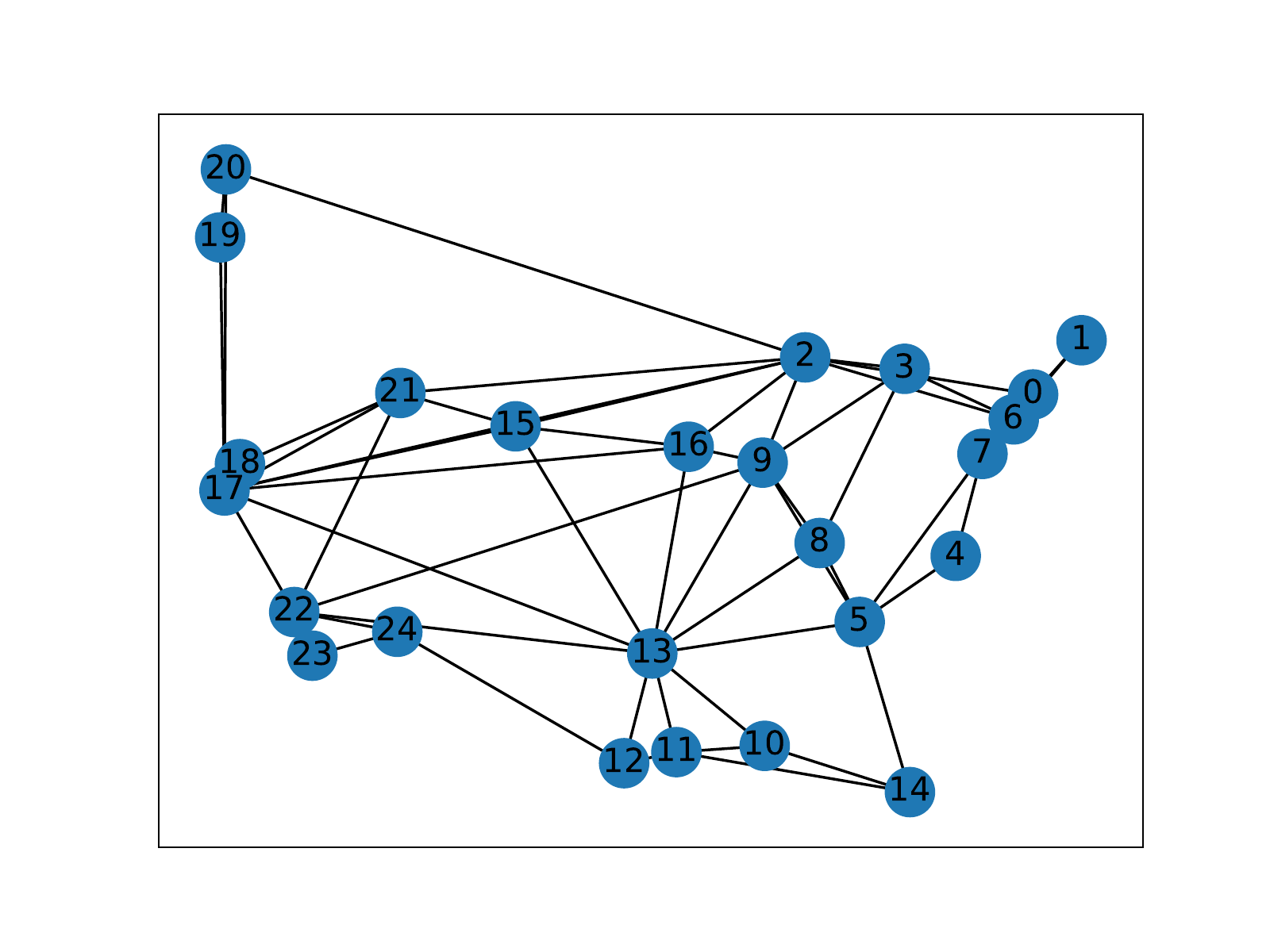}}
	\caption{AT\&T North America Topology.}
	\label{fig:topology_2}
	\vspace{-0.0cm}
\end{figure}

\begin{figure}[t]
	\centerline{\includegraphics[scale=0.6]{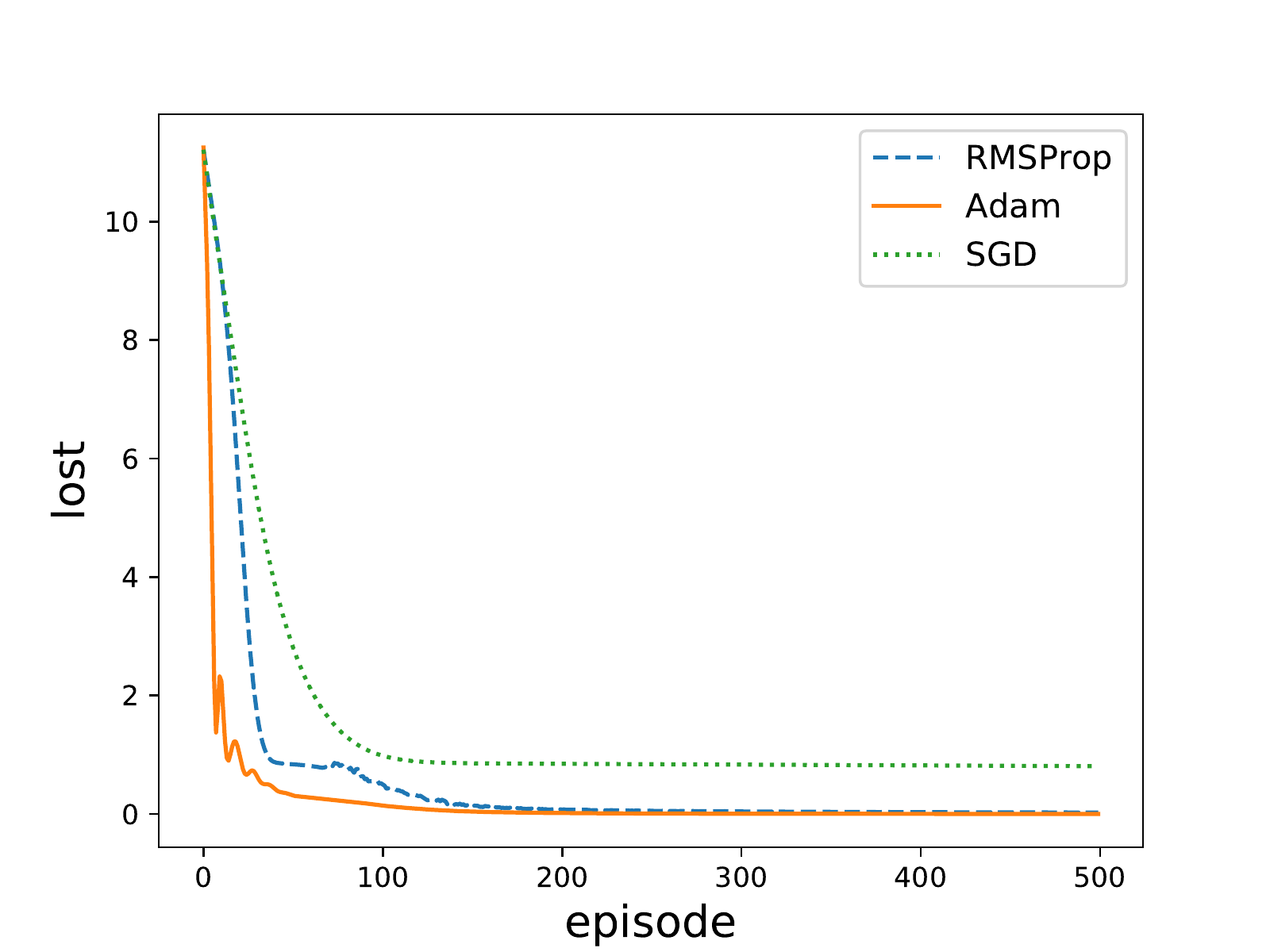}}
	\caption{Pre-Training result with different optimization algorithms.}
	\label{fig:pre-train-lost}
	\vspace{-0.0cm}
\end{figure}

\begin{figure}[t]
	\centerline{\includegraphics[scale=0.45]{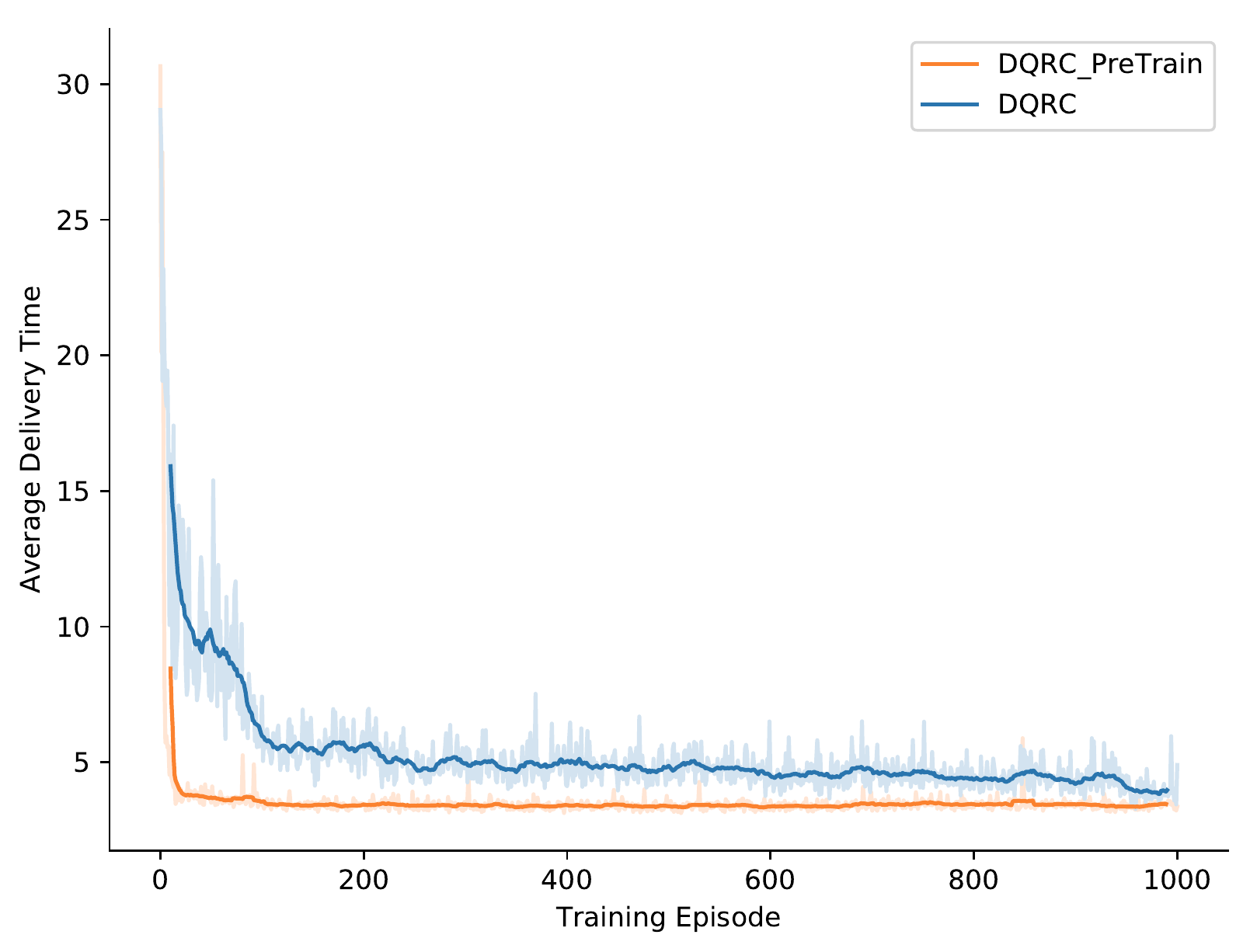}}
	\caption{Comparison between DQRC with and without pre-training.}
	\label{fig:pre-train}
	\vspace{-0.0cm}
\end{figure}

\begin{figure}[t]
	\centerline{\includegraphics[scale=0.43]{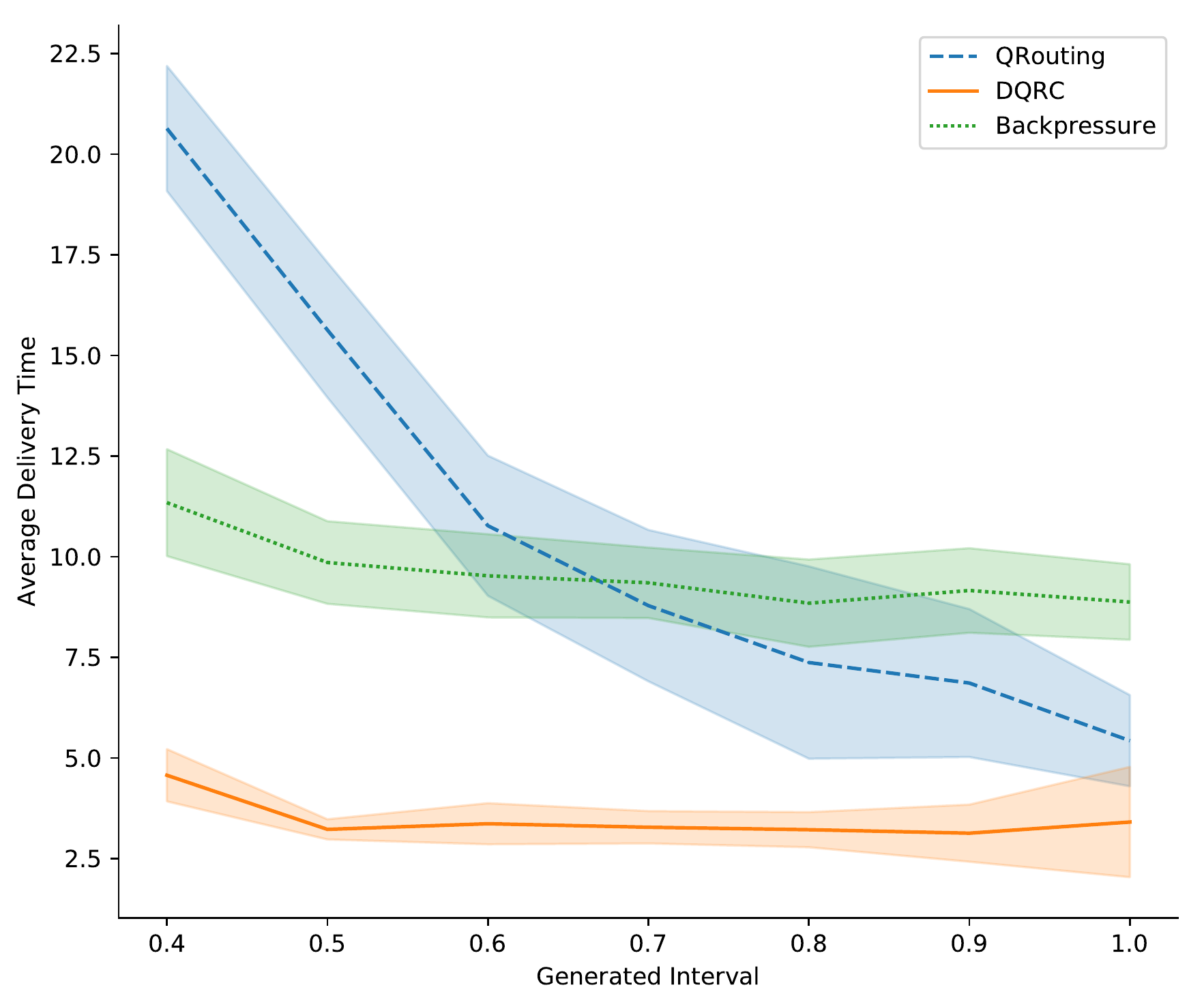}}
	\caption{Offline test with different network loads in AT\&T network.}
	\label{fig:offline_2}
	\vspace{0.0cm}
\end{figure}

\subsection{Complex Topology}
\label{sec:complex}
In the above experiments, the 3$\times$3 topology in Fig. \ref{fig:topology} is symmetrically designed and each node has no more than 4 neighbor nodes. To test the scalability of our proposed algorithm, we expand the network scale to a 25-nodes and 56-edges topology depicted in Fig. \ref{fig:topology_2}. The topology is taken from topology zoo \cite{topology_zoo} and represents real AT\&T North America network. Each vertex represents a city, for example node 0 is New York City, and each edge is a bidirectional link. The connection of routers becomes complex and each router has more choices to make, thus increasing the difficulty of decision making. The attributes of the new topology are the same as those in Section \ref{sec:se} except that node 17 is viewed as the busy ingress-router and node 8 is viewed as the busy egress-router.

In multi-agent DRL, the addition of more agents will increase the convergence difficulty of neural networks. To address this problem, we introduce the technique of pre-training. In the neural network designed in Section \ref{sec:dnn}, the value of each output neuron represents the estimated remaining delivery time after choosing the corresponding neighbor node. We take shortest-path algorithm as an initial guide to grasp the topology information. In detail, firstly we calculate the lengths of the  shortest path between each source-destination pair; Then we treat these lengths as labels and craft states by editing only the current destination and setting the other three part of input information to zeros; Finally, we perform supervised learning to train the neural network before putting it into the process of RL.

We conducted pre-training with three optimization algorithms: Stochastic Gradient Descent (SGD), Adam and RMSProp. Fig. \ref{fig:pre-train-lost} depicts the average lost change of all the agents with respect to the episode of pre-training. We can clearly see that, with Adam or RMSProp as the optimization algorithm, the average lost decrease sharply in the first few episodes, and after about 200 episodes, the neural network of each nodes converges with average lost close to zero. However, pre-training with SGD optimization converges to a higher level with slower speed. Therefore, we choose Adam as the final optimization algorithm in the pre-training process.

Fig. \ref{fig:pre-train} shows the performance gap between DQRC with and without pre-training during the RL training process. We find that, in such a complex topology with 25 nodes, DQRC finds itself takes a very long period converge with random initialization. However, after the implementation of pre-training, DQRC reaches a fairly good performance in the initial state and converges quickly to a low delay level.

\begin{figure}[t]
	\centerline{\includegraphics[scale=0.45]{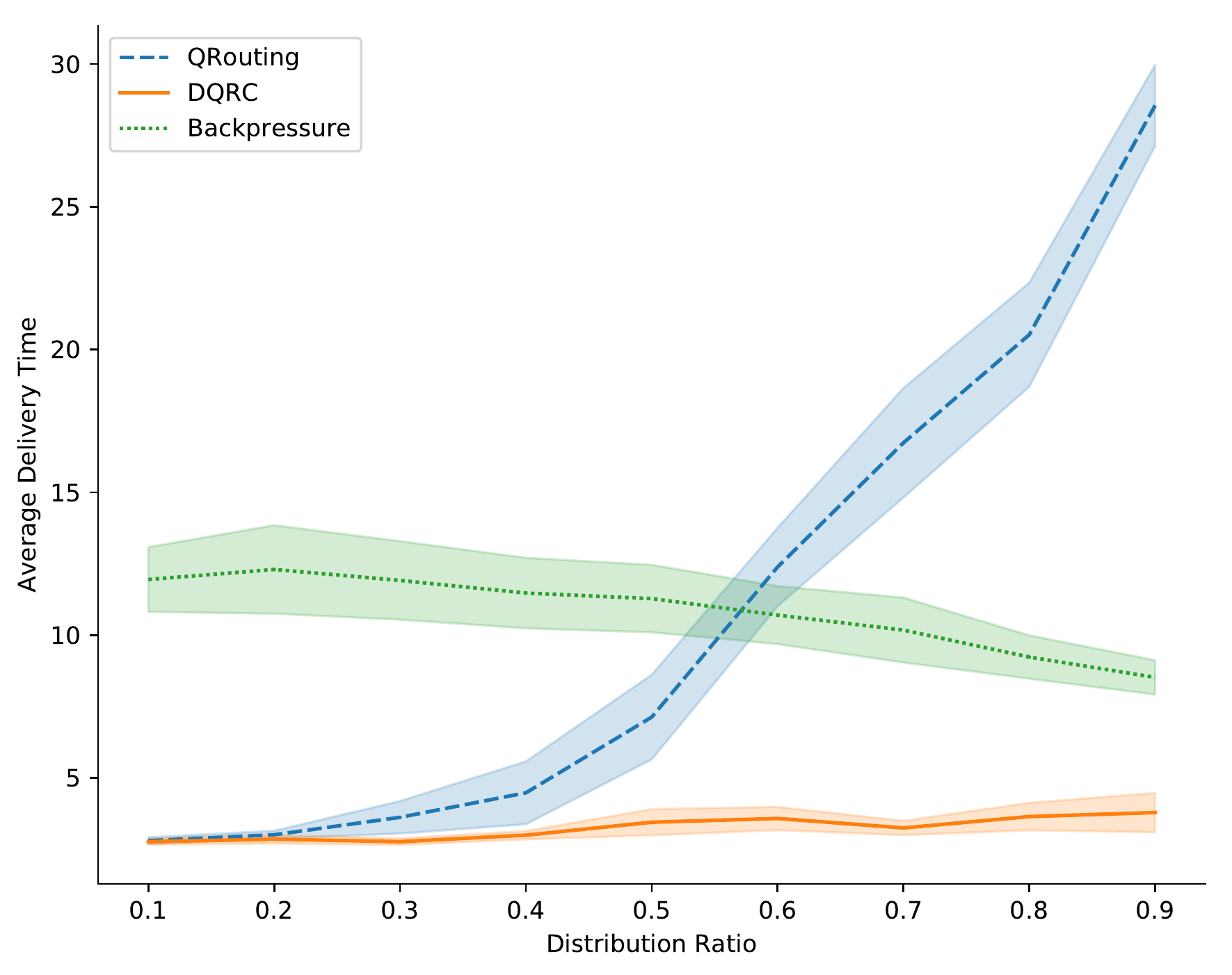}}
	\caption{Offline test with different distribution ratios in AT\&T network.}
	\label{fig:offline_ratio_2}
	\vspace{-0.0cm}
\end{figure}

In line with the experiment setting in Section \ref{sec:experiment}, we execute online-tests in two cases: (1) fixing the distribution ratio and changing the network load level with different packet generated intervals; (2) fixing the generated intervals and changing the the distribution ratio. The simulation results are illustrated in Fig. \ref{fig:offline_2} and Fig. \ref{fig:offline_ratio_2} respectively. Unsurprisingly, in this complicated topology, the variation trends of the average delivery time with respect to the network load level and distribution ratio have fairly consistency with those in the 3$\times$3 topology. The similar results in both topologies demonstrate good robustness of DQRC.

\subsection{Performance Analysis}
 With the results in both Offline-training Online-test and Online-training Online-test tests with different topologies, DQRC can be regarded as a combination of Q-routing and Backpressure, achieving not only shorter average delivery time but also greater stability in dynamically changing networks. We will clarify the main reasons for this result from the perspective of the routing policy learned by each algorithm. To simplify our interpretation, the following analysis is based on the 3x3 topology (Fig. \ref{fig:topology}).

\textbf{Why does DQRC possess the features of Q-routing and Backpressure?} 
From Fig. \ref{fig:gi} and Fig. \ref{fig:dr}, we can see that in both cases DQRC resembles Q-routing in light network load and Backpressure in heavy network load. This is an inevitable result owing to the neural network structure. In the anterior description of neural network design in Section \ref{sec:dnn}, the deep recurrent neural network of DQRC is constructed on the basis of Q-table in Q-routing, which is used to store Q-values for each state-action pair. Therefore, DQRC reserves the ability to estimate the remaining delivery time for all possible decisions. When the network is relatively light-loaded, both Q-table and neural network can give the agent an accurate estimate of Q-value to help with decision making. But this consonance will be disturbed by the rise of network load, in which case the traits of Backpressure become useful. Different from Q-routing, agents in DQRC are no longer isolated without communication with each other, and in contrast, they will exchange their queue information with adjacent agents. With this extra information, agents in DQRC can be aware of the pressure in the queue of neighbor nodes, thus avoiding conveying too many packets to busy nodes. DQRC has grasped the core idea of Backpressure which has been proved its excellent performance especially in heavy traffic pattern.

\textbf{What are the advantages of DQRC?} 
In addition to time delay estimation with neural network and information sharing between neighbor nodes, we also introduce new techniques into DQRC: (1) as described in Section \ref{sec:dnn}, we take full advantage of the representation ability of the neural network by expanding the input dimension. With the introduction of history actions and future destinations, DQRC can make more accurate estimate of the Q-value. (2) With the LSTM layer in neural network, DQRC can capture invisible connections among input elements and temporal features of sequential decisions.

Whenever the agents of Q-routing choose an action, the only information they can utilize is the destination of the current packet, leading to the same routing decision for packets with the same destination. For this reason, the sole input will cause violent fluctuations during the process of offline training (Fig. \ref{fig:training}). However, with additional information as the network input, the agents of DQRC can execute different but effective routing policy for each packet despite the same destination. More precisely, we evaluate the case where the generated interval and distribution ratio are set to 0.1ms and 70\% respectibly in online test. The network load is so heavy that a large number of packets are waiting in the queue of node 0 to be transmitted to node 8. For these packets, the agent of node 0 has two choices: sending them to node 1 or node 3. The well-trained agent of node 0 of Q-routing follows the shortest path and therefore all those packets will be sent to node 1. Under this strategy, serious congestion will occur unavoidably in the links 0$\to$1, 1$\to$2 and 1$\to$4, which eventually lead to longer delivery time. However, DQRC can overcome this difficulty cleverly. Before making decisions for every packet destined for node 8 at node 0, the agent will first collect the information about the actions the last five packets have taken and the nodes the next five packets are destined for and then check the queue length of node 1 and node 3 with communication. For example, when the last five packets were sent to node 1, the agent will decide to send the current packet to node 3 to avoid long queuing delay. Similarly, after some packets were sent to node 3, the agent will change its policy and decide to transfer the packet through node 1 again. Therefore, the great packet burden can be shared by node 1 and node 3 in DQRC instead of being born only by node 0 in Q-routing. 

Of course, it is possible that different kinds of information will lead to various decisions, so agents need to train themselves to judge which kind of information is more crucial and needs more attention. After the self-adaptation in training process, the astute agents have the ability to grasp the dynamic changes in the network and adjust their routing policies accordingly, which, shown in our test result, can gain shorter average delivery time and a better performance.

\newcommand{\tabincell}[2]{\begin{tabular}{@{}#1@{}}#2\end{tabular}} 
\begin{table}
    \centering
    \caption{Test with different number of hidden layers}
    \begin{tabular}{c c}
    \toprule
    \tabincell{c}{Number of hidden layers\\(between fc1 and LSTM layer)} & Average delivery time\\
    \midrule
    0 & 4.28\\
    1 & 4.17\\
    2 & 4.11\\
    3 & 4.33\\
    \bottomrule
    \label{tab:tab1}
    \end{tabular}

    \caption{Test with different number of neurons}
    \begin{tabular}{c c}
    \toprule
    \tabincell{c}{Number of neurons} & Average delivery time\\
    \midrule
    32 & 5.26\\
    64 & 4.96\\
    128 & 4.11\\
    256 & 4.71\\
    \bottomrule
     \label{tab:tab2}
    \end{tabular}

    \caption{Test with different intervals of information sharing}
    \begin{tabular}{c c}
    \toprule
    \tabincell{c}{Intervals of\\information sharing (ms)} & Average delivery time \\
    \midrule
    0 & 4.11 \\
    1 & 4.32 \\
    2 & 4.75 \\
    3 & 4.89 \\
    4 & 5.19 \\
    5 & 5.35 \\
    \bottomrule
    \end{tabular}
    \label{tab:tab3}
\end{table}

\subsection{Deep Dive into DQRC} 
In this part, we are committed to exploring the intrinsic disciplines within DQRC by evaluating its performance under the following situations: (1) we change the architecture of the neural network with different quantities of neurons and dense layers, (2) we modify the communication interval of information sharing among agents, (3) we rule out a certain element of DQRC including extra information, shared information and LSTM layer. Note that the following experiments are conducted with setting generated interval at 0.5s and distribution ratio at 70\%.

\textbf{Neural network architecture.} Starting with the original architecture (Fig. \ref{fig:lstm_network}), we altered a range of neural network parameters to understand their impact on the performance of DQRC. First, after fixing the number of neurons in each hidden layer at 128, we varied the number of hidden layers between the first hidden layer (fc1) and LSTM layer in the architecture of DQRC. The comparison results are shown in TABLE \ref{tab:tab1}. We find that the default setting yields the best performance, but the delay gap between it and the others is very small, showing DRQC's great robustness to the depth of neural network. Then with the default layer number, we varied the number of neurons in each hidden layer and LSTM layer. The changes of these parameters are synchronous, i.e., when 64 neurons are used, the neuron number of each sub-layer in the first hidden layer is 16. Results from this alteration are presented in TABLE \ref{tab:tab2}. We can see that the number of neurons have higher influence on DQRC's performance. Too few or too many neurons will lead to performance degradation. In conclusion, carefully-choosing hyper-parameters of neural network is needed for better estimation.

\begin{figure}[t]
	\centerline{\includegraphics[scale=0.5]{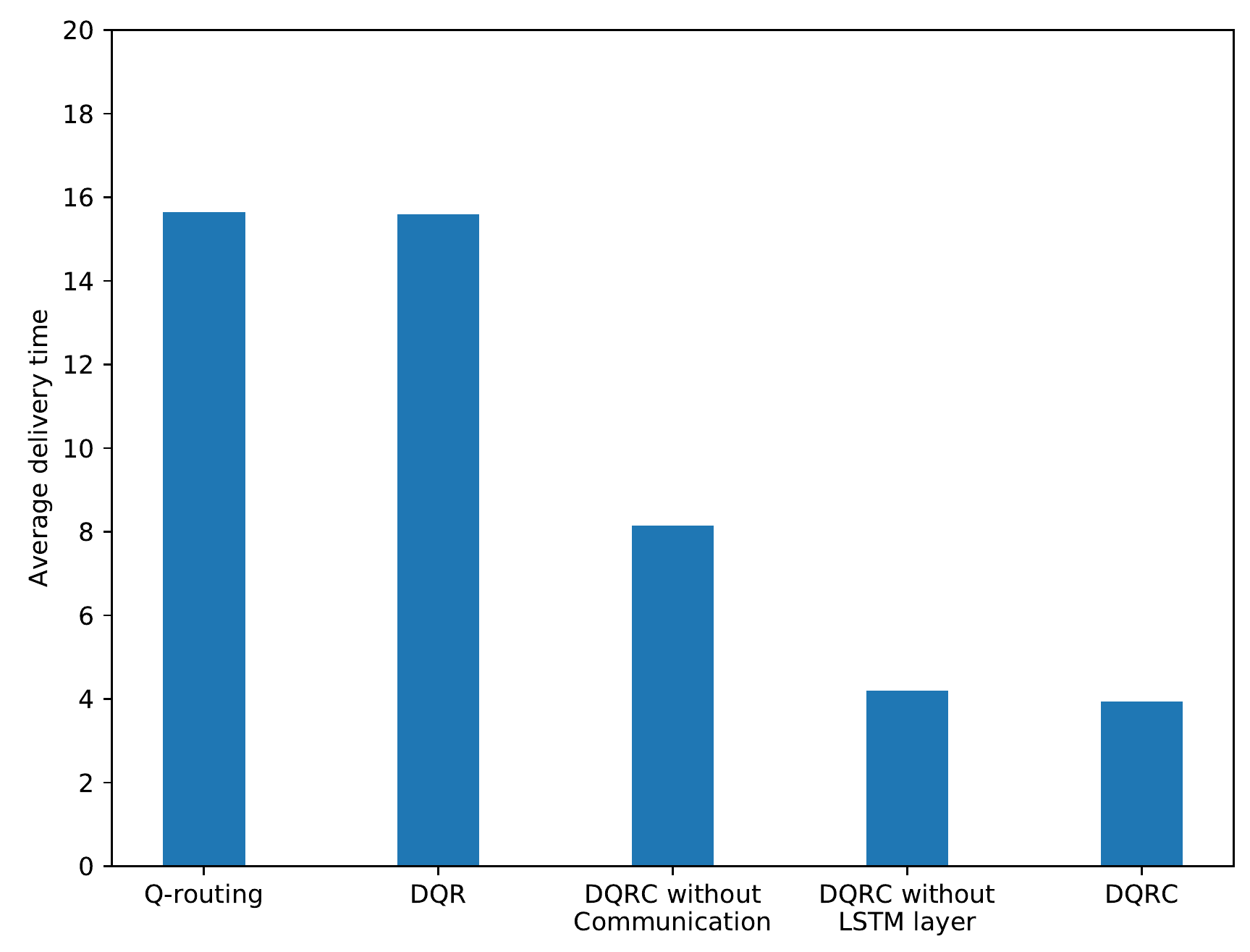}}
	\caption{Test with different algorithms.}
	\label{fig:rule_out}
	\vspace{-0.0cm}
\end{figure}

\textbf{Communication interval.} In our original design, each agent can collect queue lengths of neighbor nodes immediately whenever making a decision. However, when applying DQRC to a network with delayed feedback, the time delay during the transmission of shared information may impede the adoption of DQRC in practice. Therefore, we modified the communication interval of information sharing among agents to test its influence on DQRC's performance. With communication interval ranging from 1ms to 5ms, the test result is shown in TABLE \ref{tab:tab3}. We find that the average delivery time steadily rises as we increase the communication interval. Despite slight performance degradation, the average delivery time of DQRC is still less than half of that of Backpressure, showing DQRC's feasibility in non real-time network.

\textbf{Element in DQRC.} In order to identifies the rationality of each element in DQRC, we put forward another three algorithms on the basis of DQRC. (1) DQRC without communication: we forbid the communication process in DQRC and eliminates shared information in the state space, (2) DQRC without LSTM: we delete the LSTM layer in the original neural network (Fig. \ref{fig:lstm_network}), (3) DQR: we include only current destination into the input information. 

The comparison result with different algorithms is shown in Fig. \ref{fig:rule_out}. We can see that the lack of extra information or shared information will lead to remarkable performance degradation. If not equipped with LSTM layer, the average delivery time of DQRC will increase by 10 percent.
As for DQR and Q-routing, the only difference between these two algorithms is the representation of Q-value but the learning algorithms are identical. The neural network which is merely the approximation of Q-table would not help with the estimate of the accurate transmission time between the source and termination of packets. As a result, DQR and Q-routing who share the same input have comparable performance.

\section{Discussion}
\label{sec:discussion}
In this section, we put forward our research plan and ensuing challenges in several directions, deriving from some limitations of the current work.
 
\textbf{Other DRL algorithms.} 
The routing algorithm we propose is based on DQN \cite{DQN}, which is a classical but simple form of DRL. Thanks to the tremendous contribution researchers in the community of DRL have made, more effective DRL algorithms can be leveraged in packet routing. For example, as the optimization of the policy gradient based RL algorithm, TRPO \cite{TRPO} is combined with the neural network in continuous control domain to ensure monotonic performance \cite{bench}. Besides, based on DPG \cite{DPG}, an off-policy actor-critic algorithm DDPG \cite{DDPG} uses the neural network as a differentiable function approximator to estimate action-value function, and then updates the policy parameters in the direction of the deterministic policy gradient. 

\textbf{Realistic simulation environment.} 
In this paper, the experiments we have conducted in the simulation environment are based on some restrictive conditions, as described in Section \ref{sec:se}, which would impede the adoption of our proposed routing algorithms in the realistic network with complex traffic patterns. In the future work, we will consider a more general network setting such as routers with finite queue and packets with varied sizes. 
A couple of uncertain elements like link breakage or node failure can also be taken into account.
Furthermore, NS-3 network simulator \cite{ns3} can be utilized as the standard platform to test the performance of routing algorithms.

\textbf{Multi-agent Learning.} 
The packet routing system in our current implementation is built on the multi-agent model, where each router is treated as an independent agent and learns asynchronously. However, in multi-agent environment, the inherent non-stationarity problem \cite{non_stable} during the learning process will be magnified after the application of DNN. Besides the information-sharing mechanism stated in our paper, importance sampling and fingerprint \cite{stable} provide alternative solutions to address this problem. Moreover, witnessing the benefits of cooperative learning \cite{commu}, we will analyse the performance boost that the local coordination of DNNs (e.g., parameter sharing and memory sharing \cite{commu}) can yield in our future study.

\section{Conclusion}
\label{sec:conclusion}
We presented a fully-distributed packet routing algorithm DQRC based on multi-agent deep reinforcement learning. We designed a deep recurrent neural network with proper input to make more accurate estimation for the transmission delay. With information sharing mechanism, each agent in DQRC can learn adaptive routing policy through the interaction with environment. From the preliminary experiment results, we find that, compared with Q-routing and Backpressure, DQRC can reduce the average packet delivery time to a considerable extent in different traffic patterns and topologies.

\ifCLASSOPTIONcompsoc
  \section*{Acknowledgments}
\else
  \section*{Acknowledgment}
\fi

This work is partially supported by Natural Science Foundation of China (No. 61772139), the National Key Research and Development Program of China (No.213), Shanghai-Hong Kong Collaborative Project under Grant 18510760900 and CERNET Innovation Project NGII20170209.

\ifCLASSOPTIONcaptionsoff
  \newpage
\fi

\end{document}